\documentclass[useAMS,usenatbib,usegraphicx]{mn2e}
\usepackage{amsmath,amsfonts,amssymb}
\def\bbbone{{\mathchoice {\rm 1\mskip-4mu l} {\rm 1\mskip-4mu l} {\rm
      1\mskip-4.5mu l} {\rm 1\mskip-5mu l}}} 

\title[Magnetic Field Seeding]{Magnetic Field Seeding by Galactic Winds}

\author[Bertone et al.]
{Serena Bertone$^{1,2}$\thanks{E-mail: s.bertone@sussex.ac.uk},
Corina Vogt$^{3,2}$\thanks{E-mail: vogt@astron.nl} \&
Torsten En{\ss}lin$^{2}$\thanks{E-mail: ensslin@mpa-garching.mpg.de} \\ 
$^{1}$Astronomy Centre, University of Sussex, Brighton, BN1 9QH, United Kingdom \\
$^{2}$Max Planck Institut f\"ur Astrophysik, Karl Schwarzschild Str. 1,
85741 Garching bei M\"unchen, Germany \\
$^{3}$Stichting ASTRON, P.O. Box 2, NL--7990 AA Dwingeloo, The Netherlands}

\begin{document}

\date{Submitted to MNRAS}
\pagerange{\pageref{firstpage}--\pageref{lastpage}} \pubyear{2004}
\maketitle
\label{firstpage}

\begin{abstract}

The origin of intergalactic magnetic fields is still a mystery
and several scenarios have been proposed so far: among them,
primordial phase transitions, structure formation shocks and galactic
outflows.  In this work we investigate how efficiently galactic winds
can provide an intense and widespread ``seed'' magnetisation. This may be
used to explain the magnetic fields observed today in clusters of
galaxies and in the intergalactic medium (IGM).  We use semi--analytic
simulations of magnetised galactic winds coupled to high resolution
N--body simulations of structure formation to estimate lower and upper
limits for the fraction of the IGM which can be magnetised up to a specified level.
We find that galactic winds are able to seed a substantial fraction of the cosmic volume with magnetic fields. Most regions affected by winds have magnetic fields in the range $10^{-12} < B < 10^{-8}$ G, while higher seed fields can be obtained only rarely and in close proximity to wind--blowing galaxies. These seed fields are sufficiently intense for a moderately efficient turbulent dynamo to amplify them to the observed values.
The volume filling factor of the magnetised regions strongly depends on the efficiency of winds to load mass from the ambient medium.
However, winds never completely fill the whole Universe and pristine gas can be found in cosmic voids and regions unaffected by feedback even at $z=0$. This
means that, in principle, there might be the possibility to probe the
existence of primordial magnetic fields in such regions.

\end{abstract}



\section{Introduction}
\label{introduction}


The investigation of magnetic fields in the intergalactic medium (IGM) and in
the intracluster medium (ICM) is a new challenge for modern
cosmology. While magnetic fields have been successfully detected in
clusters of galaxies using a variety of techniques, such as Faraday
rotation measurements and observations of diffuse large--scale
synchrotron emission \citep[for recent reviews see][]{carilli, widrow,
govoni}, only a handful of pioneering observations suggest that
magnetic fields of significant intensity may be present in the
IGM\footnote{We define the intracluster medium (ICM) as the gas bound
to clusters and groups of galaxies, while we call intergalactic medium
(IGM) the diffuse gas in the Universe which is not bound to any such
structure. This distinction is clear in numerical simulations, but it
may not always be so in observations.} and no accurate measurements are
available for those regions of the Universe with densities of order of
the mean cosmic density or lower.

Cluster magnetic fields detected through Faraday rotation measurements
are of the order of a few $\umu$G. \citet{kim91} measured field
strengths of about 2 $\umu$G on scales of 10 kpc in a statistical
sample of point sources observed through the Coma cluster, and Clarke,
Kronberg and B{\"o}hringer (2001) derived field strengths of 4 to 8
$\umu$G in a sample of point sources observed in 16 low redshift
clusters.  \citet{Johnston-Hollitt} have done a similar analysis of
southern galaxy clusters and provided an independent confirmation of
the statistical signal reported by \citet{clarke}.  Magnetic field
strengths of 10 to 40 $\umu$G have been reported for the cool cores of
the Hydra A and Cygnus A clusters on scales of 3--5 kpc
(\citealt{dreher}, \citealt{taylor}).  \citet{ve2004} apply a maximum
likelihood estimator for magnetic power spectra, based on the theory
of turbulent Faraday screens \citep{ensslin03, vogt03}, to a high
quality Faraday rotation map of the north lobe of the radio source
Hydra A produced by the novel {\small PACERMAN} algorithm
\citep{dolag04, vogt04}. They find a cluster central magnetic field
strength of $7\pm 2\umu$G.

There have been few attempts to detect intergalactic magnetic fields
beyond clusters. \citet{kim} detected radio-synchrotron radiation from
an extended magnetic field in the region of the Coma supercluster, with
a field strength of 0.1 to 0.01 $\umu$G. This radio emission comes from
a region with an enhanced number density of galaxies, which might
indicate a group of galaxies in the process of merging with Coma.
\citet{bagchi} report possible radio--synchrotron evidence of
intergalactic magnetic fields in the region ZwCL 2341.1+0000 at $z\sim
0.3$, where galaxies seem to be aligned along a filament.  However, in
both cases it remains controversial whether these measurements prove the
magnetisation of the diffuse IGM or instead the state of the gas
belonging to the supercluster environment.

Theoretically, three basic mechanisms could explain the presence of
magnetic fields in the ICM and IGM \citep[for a review][]{widrow}:
i) they could have a primordial origin, ii) they were seeded in the IGM
by battery effects of shock waves or inhomogeneous radiation fields, or
iii) their existence is linked to the formation and evolution of
galaxies. A primordial origin implies that weak seed fields of about
$\sim$ 10$^{-18}$ G were produced during the early phases of the cosmic
evolution \citep[see][]{rees, kron1994}.  These seed fields must have
later been amplified by extremely efficient mechanisms, if the fields
we are to observe today were of the order of a fraction of a $\umu$G.
Alternatively, seed magnetic fields may have been produced in
protogalaxies and galaxies and subsequently ejected into the ICM and
IGM. In both cases, cosmic shear and/or dynamo effects may intervene to
amplify the seed fields.

Cosmological magnetohydrodynamic simulations of magnetic fields in
galaxy clusters have been performed by \citet{dolag}, assuming a
homogeneous seed field of 10$^{-9}$ G at $z\sim 15$.  They find that
seed fields of this intensity can be efficiently amplified to reproduce
the magnetic field strength observed in local clusters. However,
inhomogeneous seed fields can similarly lead to the right amount of
cluster magnetisation: \citet{min01a} follow the evolution of the
magnetic fields produced by Biermann--battery effects in shock waves
from the cosmic structure formation process and find that they can be
further amplified by cosmic shear flows, leading to strong cluster
fields.

Two different mechanisms can explain the transport of the magnetic seed
fields produced in galaxies into the ICM and the diffuse IGM. The
magnetic energy may be easily ejected into the IGM and ICM by jets and
radio lobes emerging from powerful radio galaxies \citep{hoyle, rees,
daly, chak}. \citet{1997ApJ...477..560E} estimated the injection of
magnetic fields by radio galaxies, and \citet{voelk00} followed the
subsequent amplification through dynamo effects, finding that it can
account quite accurately for the observed magnetic field strengths and
cosmic ray densities in the ICM. \citet{kron2001} make a step
further and estimate the magnetic flux transported directly into the
IGM.  The second mechanism is related to the ejection of magnetised
interstellar gas and supernova ejecta by galactic winds.
\citet{kron1999} consider such a scenario of magnetised outflows from
star forming galaxies and find that a substantial fraction of the
intergalactic medium could have been seeded at relatively high
redshift.  However, their estimate was based on a power--law
extrapolation of the cosmic star formation rate to high redshifts,
which is not consistent with the known peak of star formation activity
at $z \sim 1$ \citep{madau}.  \citet{deyoung} follows the amplification
of a variety of magnetic seed fields by turbulent dynamo, driven by the
motion of galaxies in the ICM, and concludes that this mechanism cannot
account for the magnetic fields observed in clusters.

Galactic winds emerging from star forming galaxies have been detected
throughout the history of the Universe (e.g. \citealt{phil},
\citealt{hlsa}, \citealt{cecil2}, \citealt{fbb}, \citealt{adelberger},
\citealt{veilleux}) and are often advocated to explain the metal
enrichment of the IGM (e.g. \citealt{aguirre}; \citealt{tv};
\citealt{serena}, BSW05 hereafter).  Although it is still
controversial if winds from large galaxies can fully account for the observed level of metal enrichment of the IGM at $z\sim 3$ \citep{ashkst}, it is clear that they do
play a role in galaxy formation and there is no doubt whether they
eject a consistent fraction of the metal and gas content of galaxies
at any redshift.

Magnetic fields are clearly observed in the interstellar medium of
galaxies \citep{beck01}, where they are believed to play an essential
role during star formation. In a few cases, magnetic fields are also
observed in the galactic material outflowing from star forming galaxies
\citep{reuter, golla,tuell}.
\citet{reuter92} estimate a
magnetic field strength of 10 $\mu$G in the material outflowing from
M82, while \citet{dahlem} find a field strength of about 7 $\mu$G in
the outflow of NGC 4666.
\citet{birk99} argue that galactic winds are able not only to advect galactic magnetic fields but also to amplify them by shear flows from Kelvin--Helmholtz instability.
If a significant fraction of the outflows are
magnetised, it is reasonable to expect that they may at least partially
contribute to the magnetisation of the ICM and IGM. The transport of
the magnetic flux by galactic outflows may occur at virtually any
redshift: at higher redshifts the main contribution may come from
protogalaxies with total masses of the order of $\gtrsim 10^6$
M$_{\sun}$, while at lower redshifts galaxies with stellar masses
comparable to the ones of local dwarf galaxies or larger may become the
dominant source of magnetic fields.


In this paper, we use high resolution simulations of galaxy
formation, which include semi--analytic prescriptions for the physics
of galactic winds, to investigate the efficiency of outflows to seed the
ICM and IGM with magnetic fields and cosmic rays. Numerical models of
galactic winds do partially explain the transport of the metals
produced in galaxies into the IGM and, since there is an analogy
between the transport of metals and the transport of magnetic fields,
we investigate the possibility that galactic winds can eject seed
magnetic fields from galaxies and transport them into the IGM and ICM.
However, we do not intend to investigate in detail all the physical
phenomena connected to the transport of the magnetic fields in our
simulations, because semi--analytic models are intrinsically not
suitable to treat the magneto--hydrodynamics involved in full
detail.

We present two different scenarios for the ejection
and transport of magnetic fields by galactic winds: a ``conservative''
scenario and an ``optimistic'' scenario. For each case, we then
analyse the effect of the efficiency of winds to affect the
results. In the first scenario, we aim to provide a ``conservative''
estimate of the global magnetisation of the Universe and we only
consider the ejection and the transport of magnetic ``seed'' fields by
winds, while amplification mechanisms are neglected. As a consequence, these results should be treated as lower limits. In the second scenario, we include a simple prescription for the amplification of the seed magnetic fields during the wind evolution and we predict a level for the magnetisation of the Universe that represents an extreme case and that can be therefore regarded as an upper limit.

Given the resolution of our simulations, we focus on the evolution of
winds emerging from galaxies with $M_{\star} \gtrsim 10^8$ M$_{\sun}$
at $z \lesssim 10$. We are interested in determining the fraction in
volume of the ICM and IGM which can be magnetised by winds and we
calculate the minimum magnetic energy and the minimum cosmic ray
energy density which can be ejected into the ICM and IGM by winds.

The paper is organised as follows. In Section \ref{sim} we describe
our set of simulations and how we implement our recipes for the
transport and amplification of the seed magnetic fields on top of the prescriptions for the evolution of galactic winds. In Section
\ref{results} we present our results for the transport of magnetic
fields by winds and we discuss the possible amplification of the seed fields by cosmic shear and turbulence.
We briefly discuss the ejection, transport and amplification of cosmic rays in Section \ref{cr} and finally we draw our conclusions in Section \ref{conclusion}.


\section{The Simulations}
\label{sim}

In this work, we use the simulations of galactic winds presented in
BSW05 and we implement new semi--analytic recipes for the
transport of the magnetic fields on top of the pre--existing scheme.
We assume a $\Lambda$CDM cosmology with matter density $\Omega_m
=0.3$, dark energy density $\Omega_{\Lambda} =0.7$, Hubble constant
$h=0.7$, primordial spectral index $n=1$ and normalisation $\sigma_8 =
0.9$.

In Subsection \ref{sim1} we briefly describe the set of high
resolution N--body simulations and the semi--analytic model for galaxy
formation. We present our prescriptions for the physics of galactic winds
in Subsection \ref{sim2} and for the evolution of magnetic fields and cosmic
rays in Subsections \ref{sim3} and \ref{sim4}, respectively.

\subsection{Formation and Evolution of Galaxies}
\label{sim1}

A set of high resolution N--body simulations, ``M3'' \citep{felix}, is used to
follow the evolution of the dark matter and the formation of structures with
time. The ``field'' simulation M3 describes an
approximately spherical region of space with diameter 52 $h^{-1}$ Mpc and
average density close to the cosmic mean. The fact that only about half the
enclosed galaxies at $z=0$ are field galaxies, while the rest are in groups and
poor clusters, makes it particularly suitable to investigate the effects of
feedback on the diffuse IGM.
The M3 simulation efficiently combines a high resolution in mass with a large
simulated volume, necessary to study the effects of feedback from
galaxies with stellar masses as low as dwarfs in their proper cosmological
context.
The number of particles in the simulated region is about $7\cdot 10^7$ and the
particle mass is $1.7\cdot 10^8 h^{-1}$ M$_{\sun}$.
The simulations were performed by \citet{felix} using the parallel treecode
{\small GADGET I} \citep{gadget}.
The dark matter evolution was followed from redshift $z=120$ down to redshift
$z=0$ and 52 simulation outputs were stored between $z=20$ and $z=0$.

The formation and evolution of galaxies is modelled with the semi--analytic
technique proposed by \citet{kauffmann} in the new implementation by
\citet{semi}. Merging trees extracted from the simulations are used to
follow the galaxy population in time, while simple prescriptions for gas
cooling, star formation and galaxy merging model the processes
involving the baryonic component of the galaxies.
At $z=3$ a total of about four hundred thousand galaxies are
identified and about three hundred and fifty thousand are present at $z=0$.
The two largest clusters have a total mass of about $10^{14} h^{-1}$
M$_{\sun}$.

\subsection{Prescriptions for Galactic Winds}
\label{sim2}

BSW05 implement new prescriptions for the long--term evolution of
winds in the semi--analytic code of \citet{semi}. Their model aims to
provide simple predictions about the effects of winds on the diffuse
IGM and, in particular, on its metal enrichment history. Here, we give
a brief qualitative description of the model, but we refer the
interested reader to BSW05 for a more detailed discussion.

The dynamics of galactic winds is modelled as a two--phase
process: a pressure--driven, adiabatic phase, followed by a
momentum--driven phase. We treat these two phases separately, according
to the different physical processes that drive the expansion of winds.
During the first phase, an over--pressurised bubble of hot gas emerges
from a star forming galaxy and evolves adiabatically until its cooling
time becomes shorter than its dynamical expansion time.
This Sedov--Taylor phase of the evolution terminates when the loss of energy
by radiation becomes substantial and the material at the outer edge of the bubble starts to cool down.
During the second phase of the wind
evolution, a thin shell of cooled material accumulates at the shock
front and the dynamics of the wind becomes dominated by the momentum
imparted by the outflowing material onto the thin shell. This second
phase is the classic momentum--driven snowplough and both the shell and
the outflowing wind material are cool.  For simplicity, we make the
assumption of spherical symmetry.

According to \citet{omk}, during the Sedov--Taylor phase of adiabatic expansion
the dynamics of the outflow is described by the equation for the conservation of
energy of a spherical bubble with energy injection at the origin:
\begin{eqnarray}\label{energy}
   \frac{d E_{\textrm{w}}}{dt}&=& \frac{1}{2}\dot{M}_{\textrm{w}} v_{\textrm{w}}
   ^2 + \varepsilon 4\pi R^2 \cdot \nonumber \\
   & & \cdot \left\{ \left[ \frac{1}{2}\rho_{\textrm{o}} v_{\textrm{o}} ^2 +
   u_{\textrm{o}} - \rho_{\textrm{o}} \frac{GM_{\textrm{H}}}{R}
   \right] \left( v_{\textrm{s}} - v_{\textrm{o}} \right) - v_{\textrm{o}} P_{\textrm{o}} \right\} ,
\end{eqnarray}
where$E_{\textrm{w}}$ is the wind energy,
$R$ and $v_{\textrm{s}}$ are the radius and the velocity of the shock, $\dot{M}_{\textrm{w}}$ and $v_{\textrm{w}}$ the mass outflow rate and the outflow velocity
of the wind, $\rho_{\textrm{o}}$, $P_{\textrm{o}}$, $v_{\textrm{o}}$
and $u_{\textrm{o}}$ the density, the pressure, the outward velocity and the internal energy of the surrounding medium, respectively.
$M_{\textrm{H}}$ is the total mass internal
to the shock radius.  The entrainment fraction $\varepsilon$ is a
parameter that defines the fraction of mass that the bubble sweeps up
while crossing the ambient medium.  This equation states that the
bubbles are powered by the material outflowing from the galaxies and
that they are slowed down by gravity and by the ram pressure and the
thermal pressure of the ambient medium.

The equation of motion for a spherically symmetric thin shell
is given by the conservation of momentum \citep{omk}:
\begin{eqnarray}\label{label}
   \frac{d}{dt}\left( m v_{\textrm{s}} \right) & = & \dot{M}_{\textrm{w}} \left( v_{\textrm{w}} - v_{\textrm{s}} \right) -
   \frac{d\phi}{dR} m - \varepsilon 4\pi R^2 P_{\textrm{o}} - \nonumber \\
   & & \varepsilon 4\pi R^2 \rho_{\textrm{o}} v_{\textrm{o}} \left( v_{\textrm{s}} - v_{\textrm{o}}\right),
\end{eqnarray}
with $m$, $R$ and $v_{\textrm{s}}$ the mass, the radius and the velocity of the
shell and $\phi$ the gravitational potential of the dark matter halo.
The first term on the right--hand side represents the
momentum injected by the starburst, the second term takes into account
the gravitational attraction of the dark matter halo and the two final
terms represent the thermal and the ram pressure of the surrounding
medium.

The conservation of mass law for a pressure driven bubble is
\begin{equation}\label{monster}
\frac{dM}{dt}=\dot{M}_{\textrm{w}} + \varepsilon 4\pi R^2 \rho_{\textrm{o}} \left( v_{\textrm{s}} - v_{\textrm{o}}\right),
\end{equation}
while for a thin shell
\begin{equation}\label{mass}
\frac{dm}{dt} = \dot{M}_{\textrm{w}} \left( 1 - \frac{v_{\textrm{s}}}{v_{\textrm{w}}}\right)
		+ \varepsilon 4\pi R^2 \rho_{\textrm{o}} \left( v_{\textrm{s}} - v_{\textrm{o}}\right).
\end{equation}

Initial conditions for the mass loss rate and the velocity of winds are taken
from \citet{shu} and calibrated to match the observations of winds both in local
starbursts and in high redshift galaxies (e.g. \citealt{martin}, \citealt{fbb}).
$\dot{M_{\textrm{w}}}$ and $v_{\textrm{w}}$ are calculated as a function of the star formation rate of the galaxy blowing the wind:
\begin{equation}\label{cond1}
\dot{M}_{\textrm{w}} = 5 \cdot \dot{M}_{\star}^{0.71} K  \textrm{ M}_{\sun}\textrm{
yr}^{-1},
\end{equation}
\begin{equation}\label{cond2}
v_{\textrm{w}} = 320 \cdot \dot{M}_{\star}^{0.145} K^{-1/2} \textrm{ km s}^{-1},
\end{equation}
where $K$ is a constant that takes into account various properties of
the interstellar medium (ISM) like, e.g., the efficiency of conduction relative to the thermal conductivity of clouds, the minimum radius of clouds in the ISM and the dimension of star--forming regions (see \citealt{shu} for a comprehensive discussion).
With this assumption, the momentum
input $\dot{M}_{\textrm{w}} v_{\textrm{w}}$ is only weakly dependent
on $K$ ($\propto K^{1/2}$), with a stronger dependence on the star
formation rate ($\propto \dot{M}_{\star}^{0.855}$). The energy input
is proportional to the star formation rate, but does not depend on $K$
and therefore on all other galaxy properties.

Both $K$ and $\varepsilon$ define the mass loading of winds, but while
$K$ determines the amount of gas entrained from the ISM of the galaxy,
$\varepsilon$ is the fraction of gas swept up by winds from the ambient
medium after blow out.  The remaining fraction of mass $1 -
\varepsilon$ is assumed to be in dense clouds which are unperturbed by
the wind.  A low value $\varepsilon < 1$ may reflect either a clumpy
ambient medium or a heavily fragmented wind, while $\varepsilon \sim 1$
describes a near--homogeneous medium, which can be entirely swept up by
the wind.
For this work we assume $K=0.5$ and we consider two
values of the entrainment fraction parameter, that is $\varepsilon =
0.1$ and $\varepsilon = 0.3$. The first value represents the behaviour
of winds with inefficient mass loading, which can easily escape from
galaxies at any redshift, while the second represents highly mass
loaded winds, whose efficiency to expand far into the IGM is
significantly reduced with respect to less mass loaded winds.  BSW05
find that the results for winds only weakly depend on $K$, while the
strongest dependence in the model is on the entrainment fraction
parameter $\varepsilon$.
Currently, there are no observational constraints on the
parameter $\varepsilon$ and our simulations are unable to constrain it
on the basis of our results. \citet{serena05} try to constrain the
parameter by comparing the statistical properties of synthetic
Ly$\alpha$ spectra with observed ones, but are unable to put any
constraint on either of the two models: both scenarios are equally
likely and none is in contrast with observations. The volume filling
factor of galactic winds is similarly not accurately constrained by
observations: published estimates vary between 0.004\% and 40\% and
in principle neither of the models we present in this work can be
excluded on the basis of observations.

When winds expand inside the haloes of galaxy groups, they sweep up
the hot and possibly metal enriched intragroup medium and they are
subject to the cluster gravitational attraction, thermal and ram
pressure forces.  Winds expanding into the IGM
entrain metal--free gas, which is assumed to have constant density and
pressure and whose velocity field is determined by the Hubble flow.
In principle, winds have to be far more energetic to escape from large dark matter haloes than to escape from field galaxies or small groups, since the forces they have to overcome are stronger. In general, winds have a higher probability to escape the haloes of isolated galaxies than to escape from groups, and it is unlikely that they can escape from large clusters, although we cannot prove this in our simulations because of the lack of massive dark matter haloes. Ram pressure plays an important role to quench outflows and its influence is particularly strong in large haloes, where it can prevent winds from blowing \citep{Kapferer}. Thermal pressure in low--redshift haloes can also slow down winds considerably, but is normally negligible with respect to the other forces at high redshifts.

\subsection{Prescriptions for Magnetic Fields}
\label{sim3}

The aim of our calculations is to describe the injection of seed
magnetic fields into the IGM by galactic winds. With our current means, this
can only be done in an approximative way. One of the main simplifications we
use in the following is the assumption that the magnetic fields in winds are dynamically negligible. This does not imply that organised magnetic
fields at the base of the wind could not play a role in launching the
outflows. However, we assume that the conversion of magnetic forces into kinetic
energy happens before the fields enter the expanding wind bubble. The remaining magnetisation evolves passively within the outflow, allowing a simplified description of its evolution, as we briefly describe below. A more detailed derivation of our equations is given in Appendix~\ref{DerMagEn}.

We define the injection of the magnetic fields ejected by galaxies into the IGM as the {\it magnetic field seeding}, since this expression literally describes the process well. Potentially, shear flows in the IGM can subsequently amplify these seed fields to the observed levels. We note that the term {\it magnetic field seeding} is often used in the literature to describe the Battery--effects, which generate magnetic fields without requiring a pre--existing seed magnetization, due to corrections to the ideal magneto--hydrodynamic equations. This initial generation of magnetic fields is believed to take place in stars, in galaxies and during the early phases of the evolution of the universe. In this work, we do not attempt to model the process, but we assume that it has happened in the past, allowing galaxies to create their own magnetic field and to amplify it to the observed level. This existing magnetic field is then ejected by the galaxy and advected in to the IGM by a galactic--scale outflow.

In our conservative scenario, we treat the seed magnetic fields analogously to
the metals and we do not process them any further: this implies that we
describe the injection of the seed fields by galaxies into the outflows and
their dilution due to the expansion of winds, but that we do not take into
account any possible mechanism that could produce amplification.  With this
assumption, we can make predictions for the injection and transport of seed
magnetic fields into the IGM and ICM, which effectively result in a lower limit
of the actual magnetisation of the Universe.

We present our prescriptions for the amplification of the seed fields in Subsection \ref{optimistic}, where we describe the build--up of our optimistic scenario.
We discuss other possible mechanisms that could further amplify the seed fields transported by winds in Subsection \ref{amplB} and we make a few consideration about the transport of cosmic rays in winds in Section \ref{cr}.

\subsubsection{Conservative Scenario}
\label{conserv}

Since in this work we want to make order of magnitude predictions, we
neglect the possible influence of the magnetic fields on the evolution
of the galactic winds and we set up the new semi--analytic
prescriptions for the magnetic fields on top of the pre--existing
scheme of BSW05.  However, this approximation turns out to be well
justified: in fact, in our calculations we find that the magnetic
energy is always several orders of magnitude smaller than the total energy in
the winds.

To simulate the transport of magnetic fields by galactic winds, we
need two main ingredients: an equation for the evolution of the
magnetic fields and suitable initial conditions.  To be consistent
with the evolution of winds set by equations (\ref{energy}) and
(\ref{monster}), we derived an equation for the evolution of the
magnetic energy in winds:
\begin{equation}\label{benergy}
   \frac{d}{dt} E_{\textrm{B}} = \dot{E}_{B_{\rm in}} - \frac{1}{3}
   \frac{\dot{V_{\textrm{w}}}}{V_{\textrm{w}}} 
   E_{\textrm{B}}.
\end{equation}
$E_{\textrm{B}}$ is the total magnetic energy in a bubble, $\dot{E}_{B_{\rm
in}}$ is the rate of magnetic energy injected by winds, $V_{\textrm{w}}$ is the
wind volume and $\dot{V_{\textrm{w}}}$ is the variation in the wind volume due
to the expansion.  The second term on the right hand side of equation
(\ref{benergy}) expresses the decrease in the magnetic energy due to the
adiabatic expansion of the bubbles and is a negative contribution.
This can be deduced from the induction equation for a frozen--in magnetic field
expanding isotropically, when the shear amplification term is neglected. A detailed derivation, which discusses the underlying assumptions, can be found in Appendix \ref{DerMagEn}. Here we give a simplified argument. The induction equation yields $B \propto 1/ R^2$ and $E_{\textrm{B}} \propto V\, B^2 \propto 1/ R$. This implies that
\begin{equation}
   \frac{d}{dt} E_{\textrm{B}} \propto - E_{\textrm{B}}
   \frac{v_{\textrm{s}}}{R} \propto \frac{1}{3}
   \frac{\dot{V_{\textrm{w}}}}{V_{\textrm{w}}}, 
\end{equation}
where the numerical factor $1/3$ reflects the assumption of no special alignment between the local field direction and the local direction of the expansion of the wind.

We assume that the magnetic energy injection rate is
\begin{equation}\label{bin}
   \dot{E}_{B_{\rm in}} =
   \varepsilon_{B_{\rm in}} \frac{\dot{M}_{\textrm{w}}}{\bar{\rho}_{\textrm{in}}} =
   \varepsilon_{B_{\rm in}} \dot{V}_{\rm in},
\end{equation}
where $\dot{V}_{\rm in}$ is the volume of the injected material,
$\bar{\rho}_{\textrm{in}}$ is the density of the gas injected into the wind,
$\varepsilon_{B_{\rm in}}$ is the magnetic energy density of the wind material
outflowing from the galaxy, and the mass outflow rate $\dot{M}_{\textrm{w}}$ is
given by equation (\ref{cond1}). Equation \ref{bin} assumes that the magnetic
field injected at the base of the wind has the strength of a typical galactic
magnetic field.  Since the production of a galactic--scale magnetic field by,
for example, an $\alpha - \Omega$ dynamo mechanism is not instantaneous, we
assume that in our simulations magnetic energy can be injected in winds only
after the host galaxy has completed a full rotation. The rotational period of
the galaxy is calculated as a function of the disk radius and circular
velocity, that is $t_{\textrm{rot}} = 2\pi R_{\textrm{g}} / v_{\textrm{c}}$.

In our model, winds make a transition between the pressure driven
phase and the momentum driven phase when their cooling time becomes
shorter than their dynamical timescale. However, when simulating the
evolution of the magnetic fields in the wind, we continue to use
equation (\ref{benergy}) for the conservation of the magnetic flux
throughout the wind evolution.  This is a reasonable approximation,
because the magnetic energy can be injected into the winds
independently of the dynamical phase of the wind and equation
(\ref{benergy}) does not critically depend on any quantity linked to a
particular phase of the wind.

A crucial quantity in Eq. (\ref{bin}) is the density
$\bar{\rho}_{\textrm{in}}$ of the gas outflowing from a galaxy at a
rate $\dot{M}_{\textrm{w}}$.  A estimate of this quantity can be
derived as a function of the mass loss rate $\dot{M}_{\textrm{w}}$ of
a wind through the entire surface area $4\pi R_{\textrm{g}}^2$ of a galaxy at the velocity $v_{\textrm{w}}$:
\begin{equation}\label{density}
\bar{\rho}_{\textrm{in}} = \frac{\dot{M}_{\textrm{w}}}{4\pi R_{\textrm{g}}^2 v_{\textrm{w}}}, 
\end{equation}
where we assume that the galactic radius is a fixed fraction of
the virial radius, that is $R_{\textrm{g}} = R_{200}/10$, as in BSW05.
The magnetic energy density $\varepsilon_{B_{\rm in}}$ associated with
the mass injected in winds can be expressed as
\begin{equation}\label{eb}
  \varepsilon_{B_{\rm in}} = \frac{B_{\rm gal}^2}{8\pi} \left(
  \frac{\bar{\rho}_{\textrm{in}}}{\bar{\rho}_{\rm ISM}} \right)^{4/3},
\end{equation}
with $B_{\rm gal}$ the galactic magnetic field and
$\bar{\rho}_{\rm ISM}$ the density of the interstellar medium (ISM) of
the galaxy. Here we assume that the magnetic field lines expand
adiabatically and isotropically when the ISM is blown out of the disk
of a galaxy and is injected into a wind. The factor in parantheses in
Eq. (\ref{eb}) represents therefore the dilution of the magnetic
fields when the density of the ejected gas drops below the density of
the ISM.
The adiabtic index $4/3$ derives from the relationship $B \propto 1/R^2$ in the
induction equation, which implies $B \propto \rho ^{2/3}$ and $B^2 \propto \rho
^{4/3}$ for isotropic expansion. The density of the ISM is given by
\begin{equation}\label{rhogal}
  \bar{\rho}_{\rm ISM}=\frac{3M_{\rm ISM}}{4\pi R_{\textrm{g}}^3},
\end{equation}
with $M_{\rm ISM}$ the mass of the ISM.

Galaxies in the local Universe have magnetic fields of order 5 $\umu$G
(e.g. \citealt{dahlem}, \citealt{chyzy2}), while interacting galaxies
can have magnetic fields as high as 30 $\umu$G (e.g. \citealt{soida},
\citealt{chyzy}, \citealt{harnett}). In our simulations, we will
express
\begin{equation}\label{bfactor}
B_{{\rm gal}} = b \cdot B,
\end{equation}
where $B = 1 \,\umu$G is a conservative reference magnetic field
value at the base of the winds and $b$ is a ``fudge'' factor to rescale our
results to less conservative choices of an initial magnetic field.

\subsubsection{Optimistic scenario}
\label{optimistic}

In this Subsection we describe our simple and purposefully optimistic prescription for the amplification of magnetic fields during the wind evolution. To set up this prescription, we consider only the amplification caused by two mechanisms directly linked to the wind evolution, namely the injection of the shocked interstellar gas into the wind and the expansion of the wind into the surroundings of the source galaxy.

The first mechanism we consider is the amplification of the magnetic fields at the base of the winds, where the outflowing material escapes the visible regions of galaxies. Here, the magnetic fields embedded in the ejected gas can be amplified through Kelvin--Helmholtz instabilities. This complex process is described in detail by \citet{birk99} and we do not intend to discuss it further. What is relevant for this work is the fact that \citet{birk99}, when simulating the process, find that the magnetic energy density increases by a factor of 3, almost independently of their set of initial conditions.
We take this into account in our simulations by scaling the magnetic field we inject in winds by the fudge factor $b = \sqrt{3}$ in Eq. (\ref{bfactor}).
In addition, we do know that observations of magnetic fields in local galaxies yield strengths of about $B_{\rm gal} \sim 6 \umu$G \citep[e.g.][]{beck01, dahlem, chyzy2}. For our optimistic scenario, we therefore set the magnetic field strength associated to the galactic ISM to be $B_{\rm gal} = 6 \umu$G.
If we incorporate this value of a more realistic galactic magnetic field in the fudge factor, together with the amplification factor of $\sqrt{3}$ at the base of the wind by Kelvin--Helmholtz instability, we obtain a total fudge factor $b=10$, which we include in our simulations.

The second mechanism we consider for the amplification of magnetic fields is the effect of shear flows inside wind cavities and bubbles during the wind expansion into the ambient medium.
To estimate how strongly the magnetic fields could be amplified, we first need to determine the characteristic timescale of this mechanism. Since motions inside winds can take place on scales of the order of the wind radius $R$, an effective timescale $\tau_{\rm eff}$ can be estimated as
\begin{equation}
  \tau_{\rm eff}^{-1} = {\eta} \, \frac{v_{\textrm{s}} - v_{\textrm{H}}}{R} = 
  \eta \left (\frac{v_{\textrm{s}}}{R} - H\left( t \right) \right),
\end{equation}
where $v_{\textrm{s}}$ is the proper wind expansion velocity, $v_{\textrm{H}} = H\left( t \right) \cdot R$ is the Hubble flow at distance $R$ from the source galaxy and $H\left( t \right)$ is the Hubble constant as a function of time. The term $v_{\textrm{s}} - v_{\textrm{H}}$ represents the peculiar velocity of a wind with respect to the Hubble flow and is the only shear motion whose contribution to the amplification of the magnetic field in wind we consider. 
$\eta$ is a field amplification efficiency factor which depends on the poorly known flow geometry inside the wind.
This amplification by shear motions directly counteracts the dilution of the magnetic fields due to the adiabatic expansion term and can be incorporated in Eq. \ref{benergy} for the evolution of the magnetic energy as a positive term:
\begin{equation}\label{above}
   \frac{d}{dt} E_{\textrm{B}} = \dot{E}_{B_{\rm in}} + \left( 
   \frac{1}{\tau_{\rm eff}} - \frac{1}{3} 
   \frac{\dot{V_{\textrm{w}}}}{V_{\textrm{w}}}\right ) E_{\textrm{B}}. 
\end{equation}

If we assume that $\eta = 1$ and we remember that the ratio $\dot{V_{\textrm{w}}} / \left( 3 V_{\textrm{w}}\right)$ can be expressed as $v_{\textrm{s}} /R$, Eq. \ref{above} simplifies to
\begin{equation}\label{last}
   \frac{d}{dt} E_{\textrm{B}} = \dot{E}_{B_{\rm in}} -
   H\left( t \right ) E_{\textrm{B}}. 
\end{equation}

In this extremely optimistic case in which $\eta = 1$, the dilution of the magnetic fields by adiabatic expansion and the amplification by shear motions cancel out, leaving behind only the term $H\left( t \right ) E_{\textrm{B}}$. This factor is small with respect to the magnetic energy injection term $\dot{E}_{B_{\rm in}}$ and it therefore represents only a small correction in Eq. \ref{last}.


\section{Results}
\label{results}

In this Section we show our results for the magnetisation of the IGM
and ICM. In Subsection \ref{bwinds} we discuss the simulated
properties of the magnetic fields in winds for the conservative
and the optimistic scenarios. In
Subsection \ref{vff} we present our results for the fraction of
magnetised IGM as a function of cosmic time and we qualitatively
describe the effect of galaxy clustering on the distribution of the
magnetic fields in our simulated region.
We present our results in terms of \emph{comoving} magnetic field strengths, as it is usually done in the community. Since our simulations produce physical estimates of the magnetic fields, we use the relation $B_{\rm comoving} = B_{\rm physical} / (1+z)^{2}$ to convert from one system of reference to the other.

In the following Subsections, all results are presented for a total of four different models, which we list here for convenience. The first two models represent our conservative scenario and correspond to two different wind efficiencies: the ``$\varepsilon =0.1$'' model represents the case of efficient winds and the ``$\varepsilon =0.3$'' model the case of less efficient winds.
There are then two additional models which represent our optimistic scenario. In the following, they will be referred to as the ``$\varepsilon =0.1 A$'' model and the ``$\varepsilon =0.3 A$'' model and they correspond to efficient and less efficient wind models respectively.

\subsection{Magnetic Fields and Cosmic Rays in Winds}
\label{bwinds}

\begin{figure*}
\includegraphics[width=\textwidth]{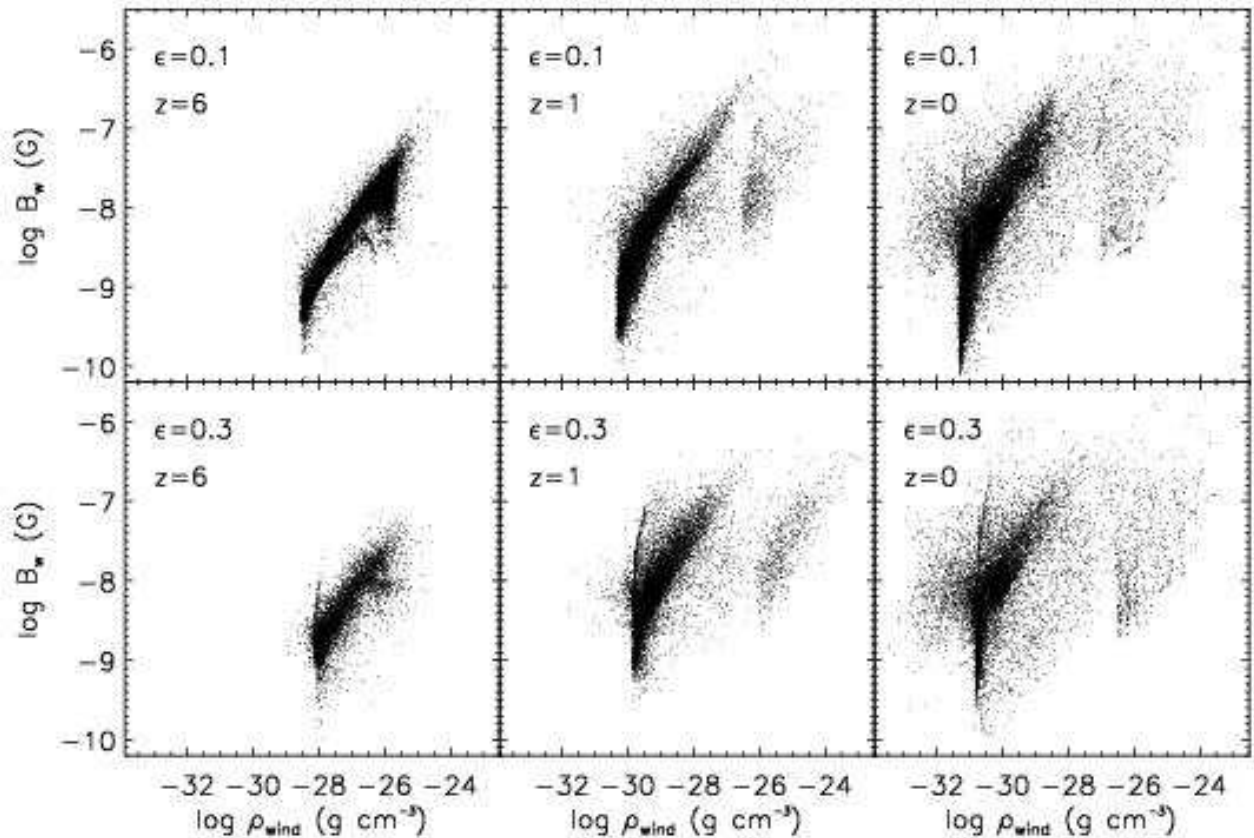}
\caption{The magnetic field strength $B_{\textrm{w}}$
for the conservative scenario in winds as a function of the wind
density for three different redshifts ($z=6$, $z=1$ and $z=0$, from
left to right) and for our two wind models with $\varepsilon =0.1$
(upper panels) and $\varepsilon =0.3$ (lower panels) in the conservative scenario.}
\label{rhob6}
\end{figure*}

Before describing our main results for the magnetisation of the
IGM, we focus briefly on the properties of the magnetic fields in
winds as they emerge from our simulations.  Our semi--analytic model
computes the total magnetic energy in a wind $E_{\textrm{B}}$, which
can easily be transformed in the root mean squared magnetic field
strength $B_{\rm w}$ by the equation:
\begin{equation}\label{conversion}
E_{\textrm{B}} = \frac{B_{\rm w} ^2}{8\pi} \cdot \frac{4\pi}{3} R^3 ,
\end{equation}
with $R$ the wind radius. $B_{\rm w}$ should
be regarded as a root mean square magnetic field strength
within the wind rather than as the value of a uniform magnetic field.

We find that $B_{\rm w}$ depends on several factors: among them, the
wind density and the total magnetic energy injected by the starburst.
Consistently, we never observe values of $B$ larger than 1 $\umu$G in our conservative scenario, which means that the dilution of the field lines due to the expansion of the wind is a significant effect at any stage of the wind
evolution.

In Fig. \ref{rhob6}, we use the two wind models in our conservative scenario to show the comoving magnetic field strength in individual winds as a function of the wind density. Results are presented for three different redshifts ($z=6$, $z=1$ and $z=0$, from left to right) and for our two wind models with $\varepsilon =0.1$ (upper panels) and $\varepsilon =0.3$ (lower panels).
Each point represents a wind at the particular snapshot time. The total
number of winds is between about $10^4$ and about $3 \cdot 10^5$,
depending on the redshift and on the model. There is a clear trend
that indicates that $B_{\rm w}$ increases with the wind density. The
scatter is mostly due to the different stages of evolution of each
wind: most winds are currently expanding into the IGM, but a number of
them are recollapsing onto the galaxies, a phenomenon known as
galactic fountain. Winds outflowing from galaxies which have only
recently assembled most of their mass may not be magnetised, since the
galaxy may not have had time to build up a magnetic field (see
Subsection \ref{sim3}).  In Fig. \ref{rhob6}, the scattered points on
the right--hand side of the distributions of magnetic fields indicate
newly formed bubbles.  These winds, which are still expanding in
proximity of galaxies, normally have higher densities and their
magnetic field strengths $B_{\rm w}$ depend more sensitively on the
magnetisation of the material outflowing from the ISM of the galaxy.
Young winds may have low $B_{\rm w}$ when little magnetic energy is
injected by the starburst because the wind mass loss rate is
small. However, in a few cases, when the wind mass loss rate is
substantial, they may also be highly magnetised (see the points in the
upper right corner of the distribution). The
points corresponding to the lowest wind densities belong either to
weakly magnetised winds or, in most cases, to winds which have joined
the Hubble flow.
By changing the fudge factor $b$ in the expression for $B_{\rm gal}$, the
results in Fig. \ref{rhob6} can be rescaled by the same factor $b$.

\begin{figure}
\includegraphics[width=8.4cm]{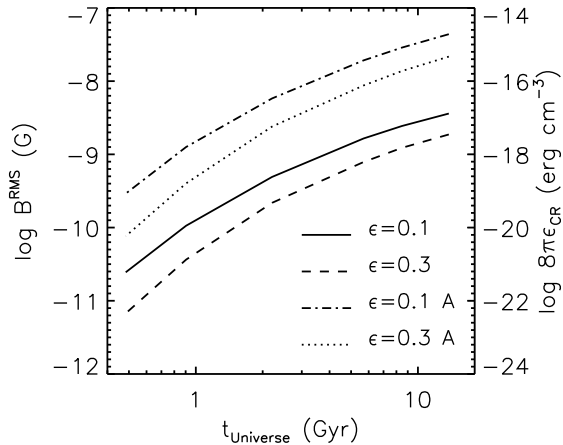}
\caption{The volume averaged RMS comoving magnetic field (left axis) and the cosmic ray energy density (right axis, see Subsection \ref{sim4})
in winds as a function of time for the two different
scenarios and the two values of our wind model parameter
$\varepsilon$. If the input parameter $B_{\rm in}$ is
modified by a fudge factor $b$, the results presented here for
the cosmic RMS magnetic field strength can be rescaled by the same factor
$b$, while the magnetic energy density can be rescaled by a factor $b^2$.}
\label{dens}
\end{figure}

In Fig. \ref{dens} we show the volume averaged RMS magnetic field strength in the Universe $B^{\rm RMS}$ (left $y$ axis) and the volume averaged cosmic ray
energy density $\epsilon_{\rm CR}$ (right $y$ axis) as a function of
the cosmic time.
The cosmic ray energy density has the same behaviour as magnetic field strength because we assume the magnetic and the cosmic ray energy densities to be in equipartition.
Results are presented for the optimistic and the conservative scenarios and for our two wind models.

The volume averaged RMS value of the cosmic magnetic fields increases with time in both scenarios and for both wind models, in agreement with the trend we observe in Fig. \ref{rhob6} for the mean value of the magnetic fields in individual winds. As expected, the magnetic field strength increases more
strongly in the optimistic scenario than in the conservative scenario: this is a consequence of the fact that in this scenario the dilution of the magnetic fields in winds because of the adiabatic expansion is largely compensated by the internal shear amplification.

Fig. \ref{dens} also shows that the level of magnetisation of the Universe depends critically on the efficiency of winds to transport matter, cosmic rays and magnetic energy out and far from galaxies.
As stated in Subsection \ref{sim2}, the winds in models with $\varepsilon = 0.1$ are much more efficient in blowing out of galaxies than in other models. As a consequence, the level of magnetisation of the Univese produced in this model is more than an order of magnitude larger than in the $\varepsilon = 0.3$ models.


\subsection{Volume Filling Factors of Magnetic Fields}
\label{vff}

In this Subsection we present our results for the volume filling
factors of magnetised regions. By volume filling factor we mean the
fraction of our simulated region affected by winds with a desired value of
the magnetic field strength. It is important to estimate
the filling factor of magnetised regions in this way, because it gives
us the possibility to make simple predictions about where the magnetic
fields may have been ejected and transported by galactic winds
and what kind of intensities we may expect them to have. Eventually,
this gives us information about the ultimate mechanism that generated
the magnetic fields we currently observe in the Universe. In the
following, we will try to answer the questions: can galactic winds
eject the seed magnetic fields that have generated the magnetic
fields in clusters and in the IGM? Or do we need other mechanisms to
explain them?

\begin{figure}
\hspace{-0.5cm}
\includegraphics[width=8.8cm]{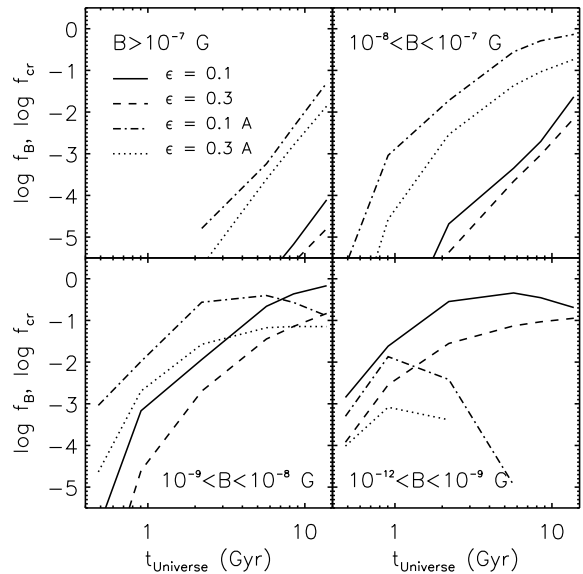}
\caption{The volume filling factors of magnetised regions as a function of the cosmic time. The four panels show the filling factors of regions with magnetic field strengths in different ranges for all four models.
Results also indicate the filling factors of the cosmic ray energy density in winds (see Subsection \ref{sim4}).
}
\label{allfb}
\end{figure}

The volume filling factor $f_{\textrm{b}}$ of magnetised regions is
calculated using the same technique used by BSW05 to obtain the volume
filling factor of galactic winds. Indeed, for each of our two models,
the sum of the filling factors of regions with different magnetisation
intensities shown in Fig. \ref{allfb} is exactly the volume filling
factor $f_{\textrm{v}}$ of winds for the two models with $(K;\varepsilon ) =
(0.5; 0.1)$ and $(K;\varepsilon ) = (0.5; 0.3)$ shown in Fig. 8 of
BSW05.
To calculate $f_{\textrm{b}}$ we superpose a grid with $512^3$ pixels
over our simulated region and we identify all those pixels which
correspond to the desired magnetic field intensity. The volume filling
factor is then simply given by the ratio between the number of pixels
with the required magnetic field strength $B$ and the total number of
pixels in the grid.  The magnetic field strength in each pixel is
recovered by summing up the magnetic energy density of each wind intersecting that particular pixel and afterwards converting it
into $B$ using equation (\ref{eb}). This conversion assumes that the
magnetic field energy is uniformly distributed inside winds:
this is only statistically correct.

Our estimates for the volume filling factors of magnetised regions as
a function of the cosmic time are shown in Fig. \ref{allfb} for all models. The four panels represent the volume filling factors of regions
with magnetic fields in four different intervals: i) $B > 10^{-7}$ G; ii) $10^{-8} < B < 10^{-7}$ G, iii) $10^{-9} < B < 10^{-8}$ G and iv) $10^{-12}
< B < 10^{-9}$ G, respectively.
Magnetic fields with strengths higher than $10^{-7}$ G have
extremely low or low filling factors at all epochs in both scenarios. On the other hand, the total filling factor of magnetic fields with
strengths in the range $10^{-12} < B < 10^{-7}$ G is already
significant at $z \sim 3$ and approaches unity at the present epoch in
models with $\varepsilon = 0.1$. This implies that in the $\varepsilon=0.1$ and in the $\varepsilon=0.1 A$ models corresponding to efficient winds, the ejected magnetic fields permeate most of the universe at $z\sim 0$.

\begin{figure}
\hspace{-0.5cm}
\includegraphics[width=8.8cm]{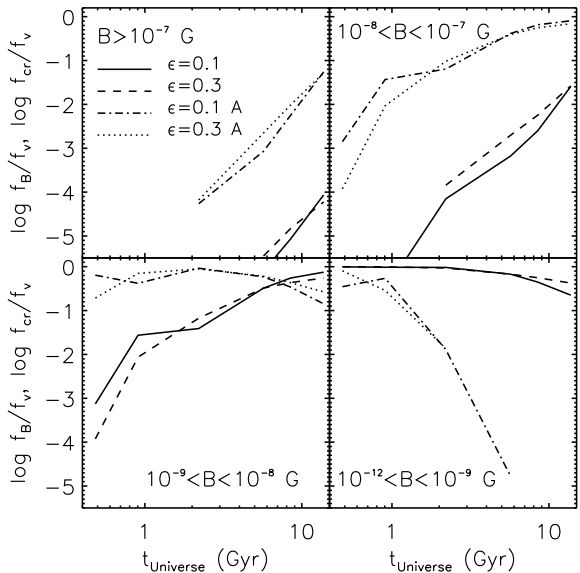}
\caption{The relative volume filling factors of magnetised regions as a function of the cosmic time. The four panels show the filling factors of regions with magnetic field strengths in different ranges for all four models.
Results also indicate the relative filling factors of the cosmic ray energy density in winds (see Subsection \ref{sim4}).
}
\label{fbfv2}
\end{figure}

In Fig. \ref{fbfv2} we show the relative volume filling factors of
magnetic fields with given strengths as a function
of cosmic time. We calculate the relative volume filling factor of
magnetised regions with strengths in the same intervals used in
Fig. \ref{allfb}. This quantity represents the ratio between the filling
factors $f_{\textrm{b}}$ shown in Fig. \ref{allfb} and the total
volume filling factor of galactic winds $f_{\textrm{v}}$ as a function of time, that is $f = f_{\textrm{b}} / f_{\textrm{v}}$. $f_{\textrm{v}}$ is the
fraction of our simulated region affected by winds, independently of
their magnetisation. Physically, $f$ indicates the fraction of space
affected by winds in which there is a magnetic field of a given
intensity. It appears that in the conservative scenario only a tiny fraction (less than 10$^{-4}$) of the volume filled by winds contains magnetic fields higher than $10^{-7}$ G, while most of the wind filled regions have
$10^{-12} < B < 10^{-8}$ G throughout the history of the Universe.
In the conservative models, most magnetised regions have a magnetic fields smaller than about $10^{-9}$ G and only at the present time most regions can reach level of magnetisations as high as $B \sim 10^{-8}$ G.

On the other hand, in the optimistic models, most winds have magnetic field strengths in the range $10^{-9} < B < 10^{-7}$. In this models, at $z=0$ most magnetised regions appear to have a magnetic field in the range $10^{-8} < B < 10^{-7}$ G and almost no regions are only weakly magnetised (that is $B < 10^{-9}$ G).
In particular, in the $\varepsilon = 0.1$ and in the $\varepsilon = 0.1 A$ models, most of the Universe is highly magnetised at $z\sim 0$. If this were the case, galactic winds would be proven to be an efficient mechanism to explain the magnetisation of filaments and the IGM observed by \citet{kim} and \citet{bagchi}, if cosmic shear or other amplification mechanisms could further amplify the seed fields by a factor of 10--1000.

\begin{figure*}
\resizebox{0.49\textwidth}{!}{\includegraphics{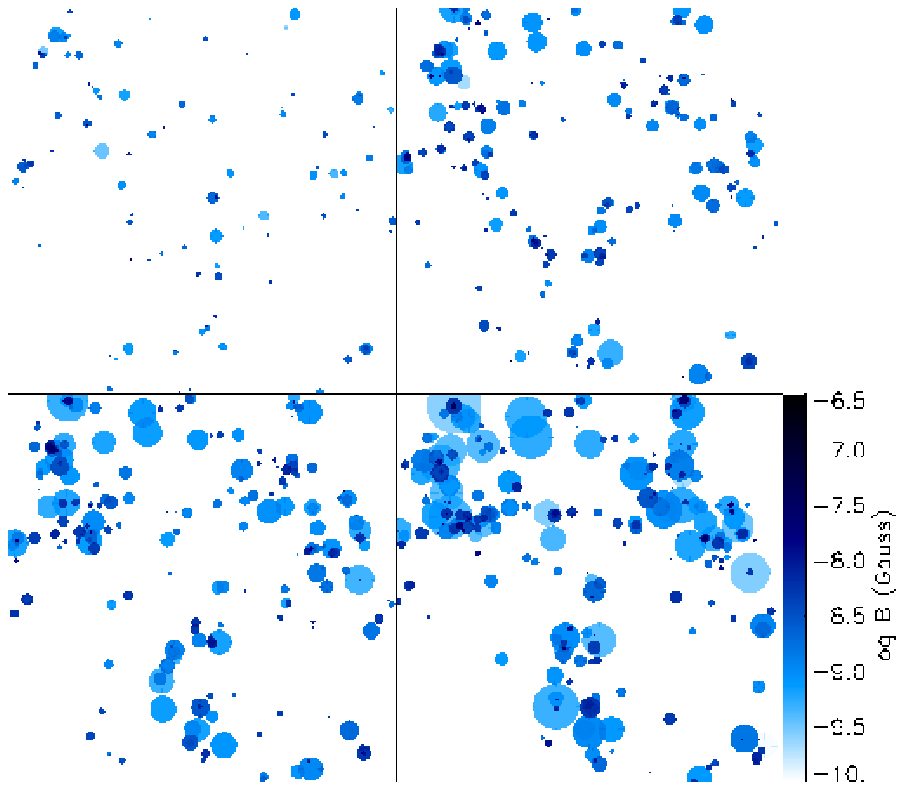}}%
\hspace{0.14cm}\resizebox{0.49\textwidth}{!}{\includegraphics{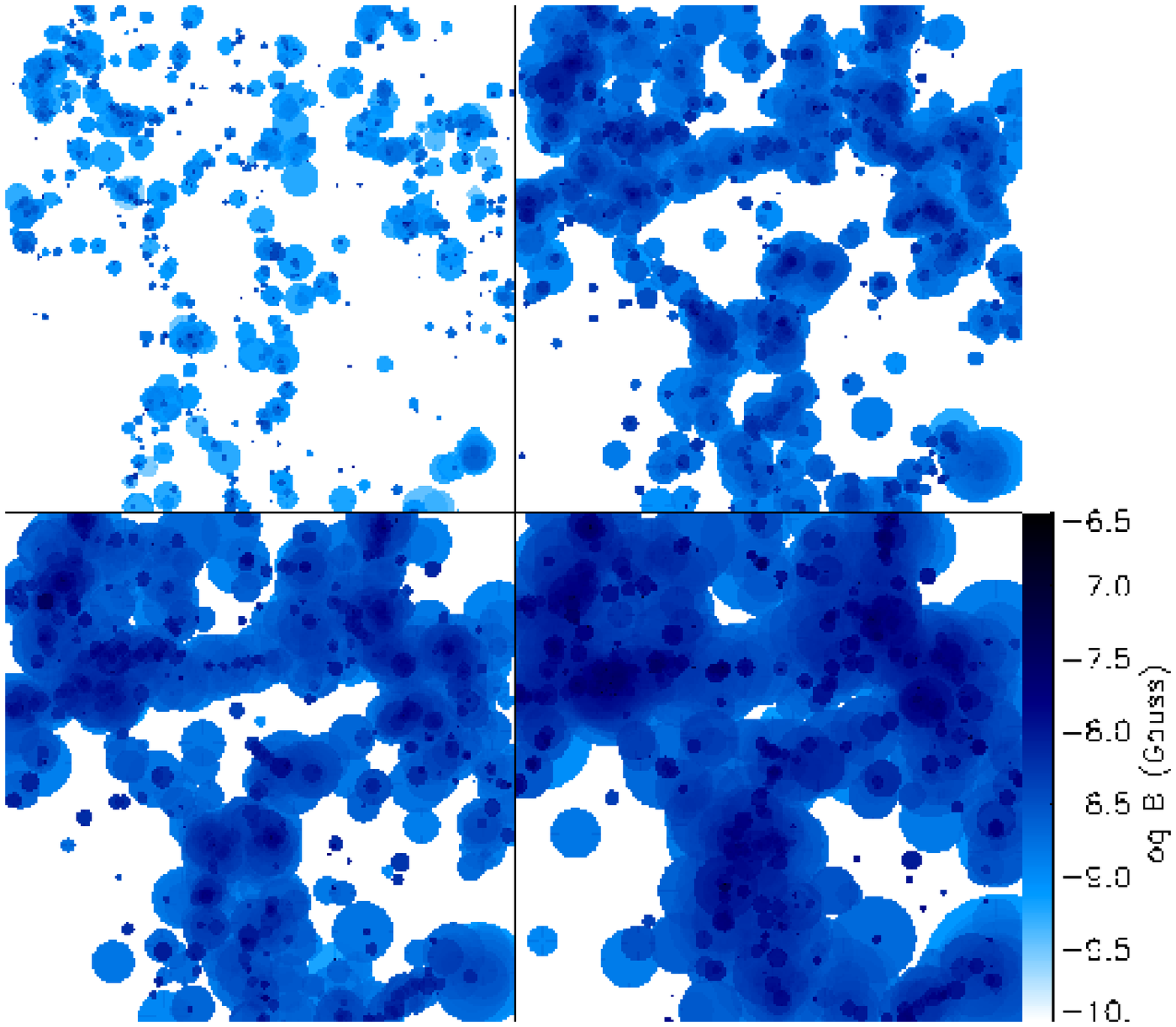}}\ \\%
\caption{The evolution of the magnetic field strength with cosmic time. Results are presented for the wind models with $\varepsilon = 0.3$ (left panels) and
with $\varepsilon = 0.1$ (right panels) of our conservative scenario. The
initial injected magnetic field is $B_{\textrm{in}} = 1$ $\umu$G.  In each
figure, from top to bottom and left to right, the redshifts of the
shown snapshots are $z = 3$, $z = 1$, $z = 0.5$ and $z = 0$
respectively.  The colour coding is the same for all snapshots and for
both models.  The comoving dimension of the simulated region is 52
$h^{-1}$ Mpc.
\label{slices}}
\end{figure*}

Fig. \ref{slices} shows slices of the magnetic field evolution in our
simulated region as a function of redshift. Results are presented for the $\varepsilon = 0.3$ (left panels) and $\varepsilon = 0.1$ (right panels) models of the conservative scenario.
In the optimistic scenario models with $\varepsilon = 0.3 A$ and $\varepsilon = 0.1 A$, results are qualitatively similar, but the colour table would have to be stretched to higher values of the magnetic field intensity.
From top to bottom and from left to right, the redshifts of the
snapshots are $z = 3$, $z = 1$, $z = 0.5$ and $z = 0$. We cut a thin
slice with a side of 52 $h^{-1}$ Mpc through our simulated region and
we colour--code the magnetic field strength from white (no field) to
black (strongest fields). The value of the magnetic field in each
pixel is the integrated value of the magnetic field $B$ recovered in
the 3D grid with $512^3$ cells used to calculate the filling factors
of magnetised regions in Subsection \ref{vff}.
There is a substantial difference between the maps of the magnetic fields in the two wind models shown here. Although the total volume filling factors $f_{\textrm{v}}$ differ by less than an order of magnitude, the effect is clearly visible in Fig.
\ref{slices}. This again is a consequence of the efficiency of
galaxies to blow winds, as determined by the parameter $\varepsilon$.
In the $\varepsilon = 0.1$ model, winds fill almost
all the simulated region, leaving us only a small chance, if any, to
observe primordial magnetic fields in the Universe at $z\sim 0$.  On the
contrary, models with higher entrainment fraction parameters and less
efficient winds as the $\varepsilon = 0.3$ model,
show that although the winds do fill a large portion of the simulated
volume, at least half or more of the intergalactic space is not
affected by feedback from galactic outflows.  This means that, in
principle and with the appropriate experiment, one could observe
primordial magnetic fields even in the present--day Universe.

The distribution of the magnetic fields in Fig. \ref{slices} shows
clearly that they are not statistically homogeneously distributed in
space, but that the magnetisation is somewhat correlated with the
underlying large scale structure of matter and galaxies. This is easily
understandable, because most of the galaxies blowing winds form in
regions of relatively high density, like filaments and groups, while fewer
objects populate very low density regions like voids. Therefore, it may
be possible that the plasma in voids is still pristine, allowing to
search for uncontaminated primordial magnetic fields.  The probability
to detect pristine gas in low density regions at $z=0$ ranges from
about 10\% for the two models with $\varepsilon = 0.1$ to about 70\% for the two models with $\varepsilon = 0.3$.

\subsection{Further amplification of the Seed Magnetic Fields}
\label{amplB}

In our simulations, most magnetised winds emerge from galaxies in
filaments or galaxies that reside in proximity of regions with
enhanced densities.  Incidentally, these regions are also those
where one would expect the amplification mechanisms such as cosmic shear
and dynamo effects to be more efficient, because of the stirring of
shear flows and turbulence by the infall of matter on the newly
forming structures.

Clusters of galaxies are expected to be only weakly rotating, if
at all, and other field amplification mechanisms than the $\alpha -
\Omega$ dynamo have to be found to explain the observed magnetic
fields.  \citet{jaffe80} and \citet{ruzmaikin} suggested that magnetic
fields could be amplified by the small scale turbulence produced by
the motion of galaxies through the ICM. However, \citet{deyoung} and
\citet{goldshmidt93} demonstrate that this mechanism alone cannot
fully produce the levels of magnetic fields observed in clusters.

There is some observational evidence for the
existence of turbulence in clusters. \citet{schuecker04} find hints of
the presence of turbulence in the Coma cluster by analysing the
pressure fluctuations revealed by X--ray
observations. \citet{ensslin05} suggest that hydrodynamical turbulence
may be responsible for the generation of the Kolmogoroff--like
magnetic power spectrum observed in the Hydra A cluster
\citep{ve2004}.

The merging of galaxy clusters is probably the most important source
of turbulence in the ICM (\citealt{tribble93}, \citealt{roettiger99b},
\citealt{norman99}).
\citet{roettiger99a} find that magnetic fields can be amplified three
times more efficiently in clusters with ongoing mergers. \citet{dolag}
and \citet{dolag02} simulate the magnetic field evolution in clusters
and find that the field strength can be increased by factors of order
$10^3$ by compression and shear of random flows during the cluster formation.
Such an amplification factor would be sufficient to bring even the seed magnetisation of a nG we find in our conservative models to the observed $\umu$G level of magnetisation in galaxy clusters.

Galactic winds emerging from galaxies in clusters can contribute
to the magnetisation of the ICM in clusters even when they do not
escape the gravitational attraction of the halo.  In fact, when star
formation occurs, some material can be shocked and ejected outside the
plane of the galaxy. This gas is then quickly stripped away from the
galaxy by the ram pressure of the hot cluster gas (\citealt{soker91},
\citealt{gaetz}) and mixed into the ICM. The stripped material would
therefore contribute to chemically enrich, shock heat and magnetise
the ICM of the cluster, although it would be unable to power a wind and reach the IGM. Unfortunately, we are currently unable to reproduce this effect
in our numerical model.

Outside clusters and bound structures, magnetic fields in winds can be
amplified by turbulence, if present, by the cosmic shear and by the compression of the outflowing gas into a thin shell, when winds become
momentum--driven and evolve like a snowplough. In this last case, the
density of the cooled material accumulating onto the shells, which is
a mixture of supernova ejecta, shocked interstellar medium and gas
accreted from the ambient medium, can become much higher than the
density of the ambient medium. Shell overdensities $\delta$ are usually in the
range $4 \bar{\rho}_{\textrm{baryons}} < \delta < 1000 \bar{\rho}_{\textrm{baryons}}$. This compression of the gas in shells
produces and analogous compression of the magnetic field lines
associated with the outflowing gas, which in turn can locally amplify
the magnetic fields. We do not attempt to model this effect in our
simulations.


\section{Extragalactic Cosmic Rays}
\label{cr}

Cosmic rays, the other non--thermal component of the intergalactic and
intracluster gas, are relativistic charged particles and, as such,
they are strongly coupled to magnetic fields. Cosmic ray electrons are
currently observed through the synchrotron radiation that they emit
when spiralling around the magnetic field lines. Since their cooling
time is dominated by the energy losses from Compton scattering and/or
synchrotron emission, energetic cosmic ray electrons cool fast,
limiting the observability of the primary cosmic ray electrons to a
short period of their lifetime.  Cosmic ray protons have not yet been
unambiguously detected, but new and up--coming gamma--ray telescopes,
such as the High Energy Gamma Ray Astronomy ({\small HEGRA}), the
Gamma Ray Large Area Space Telescope ({\small GLAST}),
the
High Energy Stereoscope System ({\small H.E.S.S.}) telescope, and the
Major Atmospheric Gamma Imaging Cherenkov Telescope ({\small
{MAGIC}}), are aimed to detect them from the gamma rays produced in
hadronic interactions with the intergalactic and intracluster gas.
The relativistic electrons and positrons generated by hadronic
proton--proton collisions are one, among others, viable explanation of
the cluster--sized diffuse radio haloes, observed in a significant
fraction of galaxy clusters \citep{1980ApJ...239L..93D, vestrand,
1999APh....12..169B, 2000A&A...362..151D, 2004A&A...413...17P}. The
amount of cosmic ray energy required to power those radio haloes is
moderate and comparable to the expected magnetic field energy
\citep{2004MNRAS.352...76P}.  Since cosmic ray protons are confined by
chaotic magnetic field lines, they are likely to accumulate in
turbulent regions of the Universe, and especially in clusters
\citep{1996SSRv...75..279V, 1997ApJ...477..560E, 1997ApJ...487..529B,
voelk00}.

\subsection{Predictions for Cosmic Rays}
\label{sim4}

In our own Galaxy, the cosmic ray energy density is of order of the
magnetic energy density \citep{beck01}. In general, the cosmic rays
that accumulate over cosmic times should be relativistic, since
sub--relativistic protons can suffer from severe Coulomb energy losses
in the dense ISM and ICM \citep[e.g.][]{1997ApJ...477..560E}. If we take
this into account, we can approximate the equation of state of the
cosmic ray gas with a polytrope with adiabatic index $\gamma_{\rm CR}
= 4/3$. This implies that the cosmic ray--producing gas suffers exactly the same
(relative) adiabatic energy losses as the (turbulent) magnetic fields.
Thus, we can assume that the cosmic ray energy injected in winds is identical to the magnetic energy density and in our wind description we can read any magnetic field energy density also as a cosmic ray energy density, that is $\varepsilon_{\rm CR} = \varepsilon_{B_{\rm in}}= B^2/ \left( 8\pi \right)$.

Our prescriptions for winds assume that the outflows are (initially) thermally driven, and not cosmic ray--driven (\citealt{dorfi04}, \citealt{bwm1991}).
We also implicitly assume that the thermal and the relativistic components of the plasma outflowing from galaxies are not coupled via Alfv\'en waves. If this were to happen, the cosmic rays would contribute momentum to the wind and an appropriate term $\dot{E}_{\textrm{CR}}$ should be included in equation \ref{label}. This injection of momentum by cosmic rays could in principle provide a wind with the necessary kick to escape from a galaxy where star formation is not very intense.

The equipartition of
energy between magnetic fields and cosmic rays may not be true in all
environments. In cosmic ray--driven outflows, for example, the cosmic
rays leave the galaxy without ejecting much gas or magnetic fields. In
this case, the contribution of winds to the cosmic ray content of the
ICM and IGM could be much higher than our estimate. Hence, our results
about the ejection of cosmic rays should be treated as lower limits.

The small energy density in injected cosmic rays can also be
substantially boosted by the action of structure formation. It is
expected that every gas volume in the IGM which ends up in the ICM is
passed by shock waves of different strength several times. Many of
these shock waves have low Mach numbers, implying that they are very
inefficient in accelerating particles out of the thermal pool. However,
any pre--existing relativistic particle population is able to gain
substantial energy by the passage of even weak shocks
\citep[e.g.][]{min01a, min01b}. Furthermore, resonant interactions with
turbulent plasma waves can also energise old cosmic ray populations to
some degree \citep[]{brun01, brun04}.  We do not attempt an
estimate of the cosmic ray amplification processes. However, it is possible that also these seed cosmic rays may have become an energetic population at present days.


\section{Conclusions}
\label{conclusion}

We have presented new semi--analytic simulations of magnetised
galactic winds emerging from star forming galaxies. Our set of N--body
simulations of structure formation in an approximately spherical
region of space of diameter 52 $h^{-1}$ Mpc combines a high mass
resolution with a region of space large enough to allow us to simulate
the evolution of the winds in their proper cosmological context.  We
have implemented new recipes for the transport of magnetic fields into
the IGM and ICM by outflows on top of the pre--existing semi--analytic
scheme for the evolution of winds of BSW05, which uses the galaxy
formation model of \citet{semi}.

The assumption that the magnetic fields only weakly affect the global evolution of winds is well justified by the finding that the total energy of the outflow is several orders of magnitude higher than its magnetic energy.
Here we have presented two different scenarios for the transport of magnetic fields in winds and for each scenario we have analysed two models corresponding to high and low wind efficiency to blow and travel far into the IGM.
As shown in \citet{serena}, the ability of winds to travel far into the IGM strongly depends on their mass loading efficieny: the more mass is accreted from the ambient medium, the lower is the probability that winds can blow out of galaxies. The mass loading efficiency is parameterized in the wind model by the constant $\varepsilon$. Unfortunately, no direct observational evidence is available to constrain this paremeter.

In both scenarios, we have neglected any possible field amplification process due to cosmic shear and turbulence in clusters. These processes are far too complex to be treated correctly in semi--analytic models.
In what we called our conservative scenario we do not consider any amplification mechanism, but we take into account the dilution of the magnetic fields due to the adiabatic expansion of the magnetic field lines inside winds. This implies that the results of our conservative models should be regarded as lower limits to any seed magnetisation of the Universe.
Conversely, in our optimistic scenario we do consider a simple prescription for the amplification of magnetic fields in winds, due to the shear motions of the material inside the wind cavities and bubbles. In the two optimistic models, we consider an efficient amplification of the magnetic fields by the internal shear motions and we set up our prescriptions by assuming that the amplification by shear motions counterbalances the dilution due to the adiabatic expansion.

Since cosmic rays can be strongly coupled to turbulent magnetic fields,
the simulations also allow us to follow the seeding
of cosmic rays by galactic winds in the IGM and ICM. Given our
assumption that the magnetic energy density equals the cosmic ray
energy density, in our simulations the cosmic rays are spatially
distributed in the same way as the magnetic fields.

We find that the magnetic fields ejected by galaxies with stellar
masses $M_{\star} \gtrsim 10^8$ M$_{\sun}$ can fill a substantial fraction
of our simulated volume, producing a mean (seed) magnetisation of
order $10^{-12}$ to $10^{-8}$ G in the conservative models and of
order $10^{-9}$ to $10^{-7}$ G in the optimistic models.
Magnetic field are not uniformly distributed in space, but rather seem to roughly follow the large scale distribution of the underlying dark matter density field.

The ejection of magnetic fields and cosmic rays by galactic winds
is efficient enough to explain the magnetic fields and cosmic ray
energy densities observed or expected in clusters, if additional
mechanisms intervene to amplify the magnetic fields and to energise the
cosmic rays by a further factor of 10 to 1000. Amplification mechanisms such as cosmic shear, dynamo effects and others (see Section \ref{amplB} for a discussion) do operate while the fields are ejected by winds into the ICM and IGM. It remains to be shown, however, whether the cosmic ray protons transported by winds can be re--accelerated by intergalactic shocks and turbulence to the level expected in galaxy clusters from the existence of radio haloes, which might be of hadronic origin.

Our model does not exclude \emph{a priori} an additional primordial
magnetic field. In fact, since at least in the high mass loading wind
scenario a non--negligible fraction of our simulated volume is
unaffected by magnetised winds, it may in principle be possible
to detect primordial magnetic fields in those regions which have not
been affected by feedback effects.


\section*{Acknowledgements}

We would like to thank Felix Stoehr for his M3 simulations
and Andrew Liddle and Simon White for reading the manuscript.
This work has been supported by the Research and Training Network ``The Physics
of the Intergalactic Medium'' set up by the European Community under contract
HPRN--CT--2000--00126.
SB was partially supported by PPARC.


\appendix
\section{Derivation of the magnetic energy evolution equation}\label{DerMagEn}

In this section we derive the equation for the evolution of the magnetic energy in winds. We introduce the set of approximations necessary to yield the
evolutionary equations used in our simulations (Equations~\ref{benergy} and
\ref{above}). This derivation should both motivate our approach and clearly show where its limitations are.

We first consider the magnetic induction equation:
\begin{equation}\label{binduction}
\frac{\partial \vec{B}}{\partial t} = \vec{\nabla} \times \vec{v} \times \vec{B}
\end{equation}
From this, we easily obtain the evolution equation of the magnetic energy
density $\varepsilon_{\textrm{B}} = B^2/(8\,\pi)$
\begin{equation}\label{depsdt}
  \frac{d }{dt}\varepsilon_{\textrm{B}} = \frac{-B^2\,\vec{\nabla} \cdot \vec{v} +
  \vec{B}\,\vec{B}\,: \vec{\nabla}\,\vec{v}}{4\,\pi}\,,
\end{equation}
where $d/dt = \partial/\partial t + \vec{v} \cdot \vec{\nabla}$  is the
total time (or convective) derivative. The magnetic energy stored in the wind
is
\begin{equation}\label{ebdef}
  E_{\textrm{B}} \equiv E_{\textrm{B}}[V_\textrm{w}] = \int_{V_\textrm{w}} \!\!\!\! d^3x \,
  \varepsilon_{\textrm{B}} = V_\textrm{w} \, \langle 
  \varepsilon_{\textrm{B}}  \rangle_{V_\textrm{w}}\,, 
\end{equation}
where $\langle \ldots \rangle_{V_\textrm{w}}$ denotes the average over the
wind volume $V_\textrm{w}$. The magnetic field strength of winds provided in
the paper should be understood in terms of this average.

We are interested in the time derivative of $E_{\textrm{B}}[V_\textrm{w}]$, for which we
need the variation of the energy density and of the volume. Since the latter is
the integration volume and therefore non-trivially differentiable, one best
introduces a coordinate system $\vec{x}(t)$, co-moving with the gas flow, which
is the solution of the equation $\dot{\vec{x}}(t) = \vec{v}(\vec{x},t)$, with
the initial condition $\vec{x}(t_0) = \vec{x_0}$. This allows a coordinate
transformation to the static coordinate system $\vec{x}_0$, so that the energy
density in a sub-volume $V$ of $V_\textrm{w}$ evolves accordingly to:
\begin{eqnarray}
  \frac{d}{dt}\,E_{\textrm{B}}[V] \!\!\! &=\!\!\! &  \frac{d}{dt}\, \int_{V_0} \!\!\!\! d^3x_0\,
  \left|\frac{\partial \vec{x}}{\partial \vec{x}_0}\right| \,
  \varepsilon_{\textrm{B}}(\vec{x}(\vec{x}_0, t), t)\nonumber\\
\!\!\! & =\!\!\! &  \int_{V} \!\!\!\! d^3x\, \frac{d}{dt}\, \varepsilon_{\textrm{B}}(\vec{x},t) + \int_{V}
  \!\!\!\!  d^3x\,\varepsilon_{\textrm{B}}(\vec{x}, t)\, \left|
  \frac{\partial \vec{x}_0}{\partial \vec{x}}\right| \, \nonumber \\
\!\!\! & \!\!\! &   
  \left( 
  \left|\frac{\partial (v_x, y, z)}{\partial \vec{x}_0}\right| +
  \left|\frac{\partial (x, v_y, z)}{\partial \vec{x}_0}\right| +
  \left|\frac{\partial (x, y, v_z)}{\partial \vec{x}_0}\right| 
  \right)
  \nonumber\\
\!\!\! & =\!\!\! &  \int_{V} \!\!\!\! d^3x\,
  \frac{d}{dt}\, \varepsilon_{\textrm{B}} + \int_{V} \!\!\!\! d^3x\,
  \varepsilon_{\textrm{B}}\,   \vec{\nabla} \cdot \vec{v}\nonumber\\
\!\!\! & =\!\!\! &  \int_{V} \frac{d^3x}{8\,\pi}\, \left(2\, \vec{B}\, \vec{B} :
  \vec{\nabla}\, \vec{v} - B^2 \, \vec{\nabla}\cdot \vec{v} \right)\,.
\end{eqnarray}

The second equation is obtained by transforming back to the original coordinate
$\vec{x}$, while the third equation can be derived by using the product rule of determinants and explicit calculation. Let us imagine that the wind volume can be split into a number $N$ of sub--volumes $V_i$: within each sub--volume the magnetic field can be assumed to be statistically homogeneous. We can therefore write
\begin{equation}
 E_{\textrm{B}}[V_\textrm{w}] = \sum_i^N \,   E_{\textrm{B}}[V_i]= \sum_i^N \,  V_i\, \varepsilon_{{\textrm{B}}_i}
\end{equation}
and introduce the rescaled field strength $\vec{b} = \vec{B}/B_i$. The time
derivative is then
\begin{equation}\label{eq:helper} 
 \frac{d}{dt}\, E_{\textrm{B}} = \sum_i^N \,  \frac{d}{dt}\,  E_{\textrm{B}}[V_i] +
 \frac{dN}{dt}\, E_{\textrm{B}}[V_{N+1}]\,,
\end{equation}
where the last term accounts for the injection of a new wind volume element
$V_{N+1}$ at the base of the wind. We identify this term with
$\dot{E}_{\textrm{B}_{\textrm{in}}} = \varepsilon_{\textrm{B}_\textrm{in}}\,\dot{V}_\textrm{in}$.

Finally, to derive the simplified Equation~\ref{benergy} for the magnetic energy evolution, some further idealised approximations are necessary:
\begin{enumerate}
  \item {\bf Vanishing cross correlation} between fluid motion and magnetic
  fields. 
  \item {\bf Statistical isotropy} of the magnetic fields, which implies $\langle \vec{b}
  \vec{b} \rangle_{V_i} = \frac{1}{3} \, \bbbone$.   
\end{enumerate}
In order to make the impact of these idealisations more transparent, we
transform Eq.~\ref{eq:helper} so that it can be split into an ideal and a
non-ideal part,
\begin{equation}
   \frac{d}{dt}\, E_{\textrm{B}} = \dot{E}_{\textrm{B}_\textrm{in}} - \frac{ E_{\textrm{B}}}{3} \,
  \frac{\dot{V}_\textrm{w}}{V_\textrm{w}} +  
\sum_i^N E_{\textrm{B}_i} \, \left[ \frac{1}{3} \,
  \left(\frac{\dot{V}_\textrm{w}}{V_\textrm{w}} - \frac{\dot{V}_i}{V_i} \right)  \langle (b^2 -1)\,
  \vec{\nabla} \cdot \vec{v} \rangle_{V_i}
  +  2 \, \langle  (\vec{b}
  \vec{b} - \frac{1}{3} \, \bbbone) : \vec{\nabla} \, \vec{v}  \rangle_{V_i}  \right]
\end{equation}
where the ideal part (including the left--hand side term and the first two right--hand side terms) is identical to Eq.~\ref{benergy}. The non--ideal part includes three contributions. The first term
accounts for deviations of the relative expansion of different regions from the
overall expansion rate. This term is probably small during the pressure--driven
phase of the wind expansion, but will lead to a larger inaccuracy of our simplified description during the momentum--driven phase, when the different regions of the wind expand in a different way. The second term accounts for correlations in the expansion rate within the sub--volumes with the magnetic field strength. For dynamically unimportant fields, we expect this term to be small. For dynamically important field strengths, we expect the magnetic structures to expand faster than the unmagnetized regions, leading to an overestimate of the field strength.
However, the basic assumption of our description is that within the
wind volume (which excludes the base of the wind) magnetic fields are
not dynamically important. These two non--ideal contributions vanish if no
correlation exists between the magnetic fields and the fluid motion both on large and small scales.
The third non--ideal contribution is due to the shearing of the magnetic fields and leads in general to a field amplification. This term vanishes if the fields are both isotropic and uncorrelated with the fluid motion. Otherwise, it should
lead to dynamo action, for which we account in Eq.~\ref{above} with a very
simplified but effective description.


\bsp
\label{lastpage}

\end{document}